\documentclass[twocolumn]{jpsj3}
\usepackage{txfonts}
\usepackage{graphicx,bm,color,braket}
\usepackage{comment}

\title{Variational Wave Function for Inhomogeneous Bose--Einstein Condensate\\ with 3/2-Body Correlations} 
\author{Wataru Kohno, Akimitsu Kirikoshi, and Takafumi Kita}

\inst{Department of Physics, Hokkaido University, Sapporo 060-0810, Japan}
\abst{
We construct a variational wave function for inhomogeneous weakly interacting Bose--Einstein condensates beyond the mean-field approximation by incorporating 3/2-body correlations.
From our numerical results calculated for a system trapped by a one-dimensional harmonic oscillator, the $3/2$-body correlations give a contribution comparable to the mean-field energy toward lowering the ground-state energy.

}

\begin{document}
\maketitle
\section{Introduction}Bose--Einstein condensates (BECs) in dilute atomic gases are created by cooling identical Bose particles in a magnetic trap\cite{Andersonetal}.
They provide unique macroscopic quantum phenomena, such as quantized vortices\cite{Matthews,Abo-Shaeer,Hodby,Haljan,Spielman} and interference effects of two BECs\cite{Andrews,Anderson1,Michael}.
In particular, a competition between interparticle interactions and inhomogeneity, which originates from trap potentials or vortices, should be considered when constructing ground states of inhomogeneous systems.
In this context, the Gross--Pitaevskii equation\cite{GP1,GP2} has been used extensively for describing the behaviors of interacting condensates.
From the microscopic viewpoint, on the other hand, noncondensates also exist even in the ground state because of the interparticle interactions.
Therefore, noncondensates as well as condensates should be incorporated to describe the interacting systems adequately.

Girardeau and Arnowitt\cite{GA} constructed a variational ground state with noncondensed particles for homogeneous systems.
They incorporated $2$-body interactions between two noncondensed particles given by $NC+NC\leftrightarrow NC+NC$ using the mean-field approximation, where $NC$ denotes a noncondensed particle.
Recently, better variational wave functions for single-component\cite{kita3/2} and multicomponent\cite{WK} systems have been constructed beyond the mean-field approximation by incorporating the dynamical $3/2$-body correlations $C+NC\leftrightarrow NC+NC$, where $C$ denotes a condensed particle.
As shown in Refs.\ \citen{kita3/2} and \citen{WK},  $3/2$-body correlations play the role of decreasing the ground-state energy, and their contributions are comparable to the mean-field contributions. 

In this paper, we extend these previous studies to inhomogeneous systems to describe the competition between interaction effects and inhomogeneity beyond the mean-field treatment.
Specifically, we construct a variational wave function for weakly interacting inhomogeneous systems and apply the formulation to a system trapped in a one-dimensional (1D) harmonic potential. 
We show that the $3/2$-body correlations cause a substantial decrease in the ground-state energy, similarly to in Refs.\ \citen{kita3/2} and \citen{WK}.

This paper is organized as follows. In Sect. 2, we outline the method of constructing the variational wave function including many-body effects.
In section 3, we outline the numerical procedures and presents results. 
In section 4, we summarize the paper. 

\section{Formulation}
We consider identical Bose particles with mass $m$ and spin 0 trapped in an external potential $V(\bm{r})$. The Hamiltonian is given by
\begin{align}
\hat{H}\equiv&\int d\bm{r}_{1} \hat{\psi}^{\dagger}(\bm{r}_{1})\hat{K}_{1} \hat{\psi}(\bm{r}_{1})\notag\\
+\frac{1}{2}&\int d\bm{r}_{1}\int d\bm{r}_{2} \hat{\psi}^{\dagger}(\bm{r}_{1}) \hat{\psi}^{\dagger}(\bm{r}_{2}) U(|\bm{r}_{1}-\bm{r}_{2}|) \hat{\psi}(\bm{r}_{2}) \hat{\psi}(\bm{r}_{1}),\label{Hami1}
\end{align}
where $\hat{\psi}$ is the boson field operator, $\hat{K}_{1}$ is defined as $\hat{K}_{1}\equiv \hat{\bm{p}}_{1}^{2}/2m+V(\bm{r}_{1})$ in terms of the momentum operator $\hat{\bm{p}}$, and $U(|\bm{r}-\bm{r}'|)$ is an interaction potential.

We expand $\hat{\psi}(\bm{r})$ in basis functions $\varphi_{q}(\bm{r})\equiv\braket{\bm{r}|q}$, which are distinguished by a set of quantum numbers $q$ and satisfy orthonormality $\braket{q|q'}=\delta_{qq'}$ and completeness $\sum_{q}\ket{q}\bra{q}=1$, as
\begin{equation}
\hat{\psi}(\bm{r})=\sum_{q}\hat{c}_{q}\varphi_{q}(\bm{r})\equiv\hat{\psi}_{\rm{c}}(\bm{r})+\hat{\psi}_{\rm{nc}}(\bm{r}),\label{decouple}
\end{equation}
where $\ket{q=0}$ ($\ket{q\neq 0}$) denotes the one-particle state of condensates (noncondensates) and $\hat{\psi}_{\rm{c}}(\bm{r})\equiv \hat{c}_{0}\varphi_{0}(\bm{r})$ [$\hat{\psi}_{\rm{nc}}(\bm{r})\equiv \sum_{q\neq0}\hat{c}_{q}\varphi_{q}(\bm{r})$] denotes the field operator for condensates (noncondensates).

Using $(\hat{c}_{q},\hat{c}^{\dagger}_{q})$, Eq.\ (\ref{Hami1}) is transformed to
\begin{align}
\hat{H}=\sum_{q_{1}q_{2}}K_{q_{2}q_{1}}\hat{c}^{\dagger}_{q_{2}}\hat{c}_{q_{1}}+\frac{1}{2}\sum_{q_{1}q_{2}q_{3}q_{4}}U_{q_{4}q_{3};q_{2}q_{1}}\hat{c}^{\dagger}_{q_{4}}\hat{c}^{\dagger}_{q_{3}}\hat{c}_{q_{2}}\hat{c}_{q_{1}}\label{Hamiltonian2}
\end{align}
with 
\begin{subequations}
\begin{align}
&K_{q_{1}q_{2}}=\int d\bm{r} \varphi^{*}_{q_{1}}(\bm{r})\left[\frac{\hat{\bm{p}}^{2}}{2m}+V(\bm{r})\right]\varphi_{q_{2}}(\bm{r}),\label{K}\\
&U_{q_{1}q_{2};q_{3}q_{4}}=\int d\bm{r}_{1} \int d\bm{r}_{2}U(|\bm{r}_{1}-\bm{r}_{2}|)\notag\\
&\times \varphi^{*}_{q_{1}}(\bm{r}_{1})\varphi^{*}_{q_{2}}(\bm{r}_{2})\varphi_{q_{3}}(\bm{r}_{2})\varphi_{q_{4}}(\bm{r}_{1}).\label{U}
\end{align}
\end{subequations}
Our aim is to construct the ground-state wave function of Eq. (\ref{Hamiltonian2}) with Eqs.\ (\ref{K}) and (\ref{U}) that describes the weakly interacting inhomogeneous BEC characterized by an external potential $V(\bm{r})$.
To carry this out, we classify $\hat{H}$ according to the number of noncondensed states involved as
\begin{equation}
\hat{H}=\hat{H}_{0}+\hat{H}_{\frac{1}{2}}+\hat{H}_{1}+\hat{H}_{\frac{3}{2}}+\hat{H}_{2},
\end{equation}
where
\begin{subequations}
\begin{align}
\hat{H}_{0}=&K_{0 0}\hat{c}^{\dagger}_{0}\hat{c}_{0}+\frac{1}{2}U_{00;00}\hat{c}^{\dagger}_{0}\hat{c}^{\dagger}_{0}\hat{c}_{0}\hat{c}_{0},\\
\hat{H}_{\frac{1}{2}}=&\sum_{q_{1}\neq0}\Big(K_{0q_{1}}\hat{c}^{\dagger}_{0}\hat{c}_{q_{1}}+{\rm{H.C.}}\Big)\notag\\
+\sum_{q_{1}\neq0}&\Big(U_{q_{1}0;00}\hat{c}^{\dagger}_{q_{1}}\hat{c}^{\dagger}_{0}\hat{c}_{0}\hat{c}_{0}+{\rm{H.C.}}\Big),\\
\hat{H}_{1}=&\sum_{q_{1},q_{2}\neq0}K_{q_{2} q_{1}}\hat{c}^{\dagger}_{q_{2}}\hat{c}_{q_{1}}\notag\\
&+\sum_{q_{1}q_{2}\neq0}\Big(U_{q_{2}0;q_{1}0}+U_{q_{2}0;0q_{1}}\Big)\hat{c}^{\dagger}_{0}\hat{c}_{0}\hat{c}^{\dagger}_{q_{2}}\hat{c}_{q_{1}}\notag\\
+\frac{1}{2}&\sum_{q_{1}q_{2}\neq0}\Big(U_{00;q_{2}q_{1}}\hat{c}^{\dagger}_{0}\hat{c}^{\dagger}_{0}\hat{c}_{q_{2}}\hat{c}_{q_{1 }}+{\rm{H.C.}}\Big),\\
\hat{H}_{\frac{3}{2}}=&\sum_{q_{1}q_{2}q_{3}\neq0}\Big(U_{0q_{3};q_{2}q_{1}}\hat{c}^{\dagger}_{0}\hat{c}^{\dagger}_{q_{3}}\hat{c}_{q_{2}}\hat{c}_{q_{1}}+{\rm{H.C.}}\Big),\\
\hat{H}_{2}=&\frac{1}{2}\sum_{q_{1}q_{2}q_{3}q_{4}\neq0}U_{q_{4}q_{3};q_{2}q_{1}}\hat{c}^{\dagger}_{q_{4}}\hat{c}^{\dagger}_{q_{3}}\hat{c}_{q_{2}}\hat{c}_{q_{1}},
\end{align}
\end{subequations}
where H.C. denotes the Hermitian conjugate.

Next, we introduce the number-conserving creation-annihilation operators{{\cite{kita3/2,WK,GAN}}}. To carry this out, we give the orthonormal basis function for $q=0$ as
\begin{equation}
\ket{n}_{0}\equiv \frac{(\hat{c}^{\dagger}_{0})^{n}}{\sqrt{n!}}\ket{0} \ (n=0,1,2,\cdots,N).
\end{equation}
The ground state without correlations is given by $\ket{N}_{0}$.
The number-conserving operators are introduced as $(\hat{\beta}^{\dagger}_{0},\hat{\beta}_{0})$ for $n\geq0$ by $\hat{\beta}^{\dagger}_{0}\ket{n}_{0}\equiv\ket{n+1}_{0}$ and $\hat{\beta}_{0}\ket{n+1}_{0}=\ket{n}_{0}$
with $\hat{\beta}_{0}\ket{0}\equiv0$. These operators are expressible in terms of $(\hat{c}^{\dagger}_{0},\hat{c}_{0})$ as
 \begin{subequations}
\begin{align}
\hat{\beta}_{0}^{\dagger}&= \hat{c}^{\dagger}_{0}(1+\hat{c}_{0}^{\dagger}\hat{c}_{0})^{-\frac{1}{2}},\\
\hat{\beta}_{0}&= (1+\hat{c}_{0}^{\dagger}\hat{c}_{0})^{-\frac{1}{2}}\hat{c}_{0},
\end{align}
\end{subequations}
 and obey $(\hat{\beta}_{0}^{\dagger})^{\nu}\hat{\beta}_{0}^{\nu}\ket{n}_{0}=\hat{\beta}_{0}^{\nu}(\hat{\beta}_{0}^{\dagger})^{\nu}\ket{n}_{0}=0$ for integer $\nu\leq n$ and
$(\hat{\beta}_{0}^{\dagger})^{\nu}\hat{\beta}_{0}^{\nu}\ket{n}_{0}=0$ for $\nu>n$. Therefore, $(\hat{\beta}_{0}^{\dagger})^{\nu}\hat{\beta}_{0}^{\nu}=1$ and $\hat{\beta}_{0}^{\nu}(\hat{\beta}_{0}^{\dagger})^{\nu}\simeq1$  for $\nu=1,2,\cdots$. The latter approximation for $\nu\ll N$ is almost exact in the weak-coupling regime where the ground state is composed of the kets $\ket{n}_{0}$ with $n=O(N)$.

As a first step to construct the ground state of an inhomogeneous BEC, we give an inhomogeneous extension of the Girardeau--Arnowitt wave function\cite{GA} (GA  wave function).
First, we define the pair-correlation function as
\begin{equation}
\hat{\pi}^{\dagger}=\frac{1}{2}\sum_{q_{1}q_{2}\neq0}\phi_{q_{1}q_{2}}\hat{c}^{\dagger}_{q_{1}}\hat{c}^{\dagger}_{q_{2}}\hat{\beta}^{2}_{0},
\end{equation}
where $\phi_{qq'}=\phi_{q'q}$ is a variational parameter that characterizes the
pair excitation of particles $q$ and $q'$ from condensates. 
Using $\hat{\pi}^{\dagger}$, we introduce the ground-state wave function as
\begin{equation}
\ket{\Phi_{\rm{GA}}}=A^{-1/2}_{\rm{GA}}\exp(\hat{\pi}^{\dagger})\ket{N}_{0}=A^{-1/2}_{\rm{GA}}\sum_{\nu=0}^{[N/2]}\frac{(\hat{\pi}^{\dagger})^{\nu}}{\nu!}\ket{N-2\nu}_{0},
\end{equation}
where $[N/2]$ denotes the largest integer that does not exceed $N/2$ and $A_{\rm{GA}}$ is a normalization constant determined by $\braket{\Phi_{\rm{GA}}|\Phi_{\rm{GA}}}=1$.

$\ket{\Phi_{\rm{GA}}}$ is the vacuum state characterized by $\hat{\gamma}_{q}\ket{\Phi_{\rm{GA}}}=0$, where $\hat{\gamma}_{q}$ is the number-conserving quasiparticle operator defined as
\begin{align}
&\hat{\gamma}_{q}\equiv \sum_{q_{1}\neq0}\left(u_{qq_{1}}\hat{c}_{q_{1}}\hat{\beta}^{\dagger}_{0}-v_{qq_{1}}\hat{c}^{\dagger}_{q_{1}}\hat{\beta}_{0}\right).
\end{align}
Here, we require that $(\hat{\gamma}^{\dagger}_{q},\hat{\gamma}_{q})$ obey the Bose commutator relation.
In this case, matrices $\underline{u}\equiv(u_{q_{1}q_{2}})$ and $\underline{v}\equiv(v_{q_{1}q_{2}})$ are given in terms of $\underline{\phi}\equiv(\phi_{q_{1}q_{2}})$ and the unit matrix $\underline{1}\equiv(\delta_{q_{1}q_{2}})$ by
\begin{equation}
\underline{u}\equiv (\underline{1}-\underline{\phi}\ \underline{\phi}^{\dagger})^{-\frac{1}{2}} , \ 
\underline{v}\equiv (\underline{1}-\underline{\phi}\ \underline{\phi}^{\dagger})^{-\frac{1}{2}}\underline{\phi}.\label{uvphi}
\end{equation}
Therefore, they satisfy
\begin{equation}
\underline{u}^{\dagger}=\underline{u}, \ \underline{v}^{\rm{T}}=\underline{v}, \
 \underline{u} \ \underline{u}^{\dagger}-\underline{v} \ \underline{v}^{\dagger}=\underline{1}, \  \underline{u} \ \underline{v}=\underline{v} \ \underline{u}^{*}, \label{uv}
\end{equation}
where $\rm{T}$ denotes the transposition of a matrix.
The third and fourth relations are summarized as the following matrix form:
\begin{align}
\begin{bmatrix}
\underline{u} & \underline{v}\\
\underline{v}^{*} &\underline{u}^{*}
\end{bmatrix}
\begin{bmatrix}
\underline{u} & -\underline{v}\\
-\underline{v}^{*} & \underline{u}^{*}
\end{bmatrix}=
\begin{bmatrix}
\underline{1} & \underline{0}\\
\underline{0} & \underline{1}
\end{bmatrix}.\label{orthnormal}
\end{align}

Using Eq.\ (\ref{orthnormal}), $(\hat{c}_{q},\hat{c}^{\dagger}_{q})$ are also expressible in terms of $(\hat{\gamma}_{q},\hat{\gamma}^{\dagger}_{q})$ as follows:
\begin{subequations}
\begin{align}
&\hat{c}_{q}\hat{\beta}^{\dagger}_{0}= \sum_{q_{1}\neq0}\left(u_{qq_{1}}\hat{\gamma}_{q_{1}}+v_{qq_{1}}\hat{\gamma}^{\dagger}_{q_{1}}\right),\\
&\hat{c}^{\dagger}_{q}\hat{\beta}_{0}= \sum_{q_{1}\neq0}\left(u^{*}_{qq_{1}}\hat{\gamma}^{\dagger}_{q_{1}}+v^{*}_{qq_{1}}\hat{\gamma}_{q_{1}}\right).
\end{align}
\end{subequations}
Note that $\ket{\Phi_{\rm{GA}}}$ only includes pair processes via $\underline{\phi}$, meaning that it has no contributions from $\hat{H}_{\frac{3}{2}}$ and $\hat{H}_{\frac{1}{2}}$, i.e., $\bra{\Phi_{\rm{GA}}}\hat{H}_{\frac{3}{2}}\ket{\Phi_{\rm{GA}}}=\bra{\Phi_{\rm{GA}}}\hat{H}_{\frac{1}{2}}\ket{\Phi_{\rm{GA}}}=0$.
To incorporate $3/2$-body correlations, we need to characterize them by introducing the corresponding variational parameters as outlined below.

Next, we improve $\ket{\Phi_{\rm{GA}}}$ so that $\hat{H}_{\frac{3}{2}}$ yields finite contributions to lower the variational ground-state energy further.
The ground state with a new operator may be introduced as
\begin{subequations} 
\begin{align}
&\ket{\Phi}\equiv A^{-1/2}_{3} \rm{exp}\left(\hat{\pi}^{\dagger}_{3}\right)\ket{\Phi_{\rm{GA}}},\\
&\hat{\pi}^{\dagger}_{3}\equiv \frac{1}{3!}\sum_{q_{1}q_{2}q_{3}\neq 0}w_{q_{1}q_{2}q_{3}}\hat{\gamma}^{\dagger}_{q_{1}}\hat{\gamma}^{\dagger}_{q_{2}}\hat{\gamma}^{\dagger}_{q_{3}},
\end{align} 
\end{subequations}
where $w_{q_{1}q_{2}q_{3}}$ is a variational parameter characterized by 3/2-body correlations satisfying $\hat{P}_{q_{1}q_{2}q_{3}}w_{q_{1}q_{2}q_{3}}=w_{q_{1}q_{2}q_{3}}$ for any permutation $\hat{P}$ with three elements $(q_{1},q_{2},q_{3})$ and $A_{3}$ is the normalization constant expressed as 
\begin{align}
A_{3}&=\bra{\Phi_{\rm{GA}}}\rm{exp}\left(\hat{\pi}_{3}\right)\rm{exp}\left(\hat{\pi}^{\dagger}_{3}\right)\ket{\Phi_{\rm{GA}}}\notag\\
&= \exp\left(\frac{1}{3!}\sum_{q_{1}q_{2}q_{3}\neq0}|w_{q_{1}q_{2}q_{3}}|^{2}+O\Big(|w|^{4}\Big)\right). \ \label{lnA1}
\end{align}
Here, we omit the higher-order terms $O\Big(|w|^{4}\Big)$ in the present weak-coupling consideration.
In this case, we obtain
\begin{equation}
\bra{\Phi}\hat{\gamma}^{\dagger}_{q_{1}}\hat{\gamma}^{\dagger}_{q_{2}}\hat{\gamma}^{\dagger}_{q_{3}}  \ket{\Phi}=\frac{\delta {\rm{ln}}A_{3}}{\delta w_{q_{1}q_{2}q_{3}}} \simeq w^{*}_{q_{1}q_{2}q_{3}}.\label{lnA2}
\end{equation}
Note that $\bra{\Phi}\hat{\gamma}^{\dagger}_{q_{1}} \hat{\gamma}_{q_{2}}\hat{\gamma}_{q_{3}}\ket{\Phi}$, $\bra{\Phi} \hat{\gamma}_{q_{1}}\hat{\gamma}_{q_{2}}\ket{\Phi}$, and their complex conjugates are neglected since they are all higher-order contributions. In addition, we have $\bra{\Phi}\hat{H}_{\frac{1}{2}} \ket{\Phi}=0$.

On the basis of $\ket{\Phi}$, we obtain expressions for the ground-state energy and self-consistent equations embodying energy-minimum conditions.
To express the ground-state energy explicitly, we define the following quantities:
\begin{subequations}
\begin{align}
&\rho_{q_{1}q_{2}}
\equiv\bra{\Phi} \hat{c}^{\dagger}_{q_{2}}\hat{c}_{q_{1}} \ket{\Phi}=\rho^{*}_{q_{2}q_{1}}\notag\\
&=\frac{1}{2}\sum_{q_{3}\neq0}\Big(u_{q_{1}q_{3}}u^{*}_{q_{2}q_{3}}+v_{q_{1}q_{3}}v^{*}_{q_{3}q_{2}}\Big)\notag\\
&+\sum_{q_{3}q_{4}\neq0}\Big(u_{q_{1}q_{3}}u^{*}_{q_{2}q_{4}}+v_{q_{1}q_{4}}v^{*}_{q_{2}q_{3}}\Big)\rho^{\frac{3}{2}}_{q_{3}q_{4}}-\frac{1}{2}\delta_{q_{1}q_{2}}, \\
&F_{q_{1}q_{2}}
\equiv\bra{\Phi} \hat{c}_{q_{1}}\hat{c}_{q_{2}}(\hat{\beta}^{\dagger}_{0})^{2} \ket{\Phi}
=F_{q_{2}q_{1}}\notag\\
&=\sum_{q_{3}\neq0}u_{q_{1}q_{3}}v_{q_{3}q_{2}}
+\sum_{q_{3}q_{4}\neq0}\Big(u_{q_{1}q_{3}}v_{q_{2}q_{4}}+v_{q_{1}q_{4}}u_{q_{2}q_{3}}\Big)\rho^{\frac{3}{2}}_{q_{3}q_{4}},\\
&W_{q_{1}q_{2};q_{3}}
\equiv\bra{\Phi} \hat{c}^{\dagger}_{q_{3}}\hat{c}_{q_{2}}\hat{c}_{q_{1}} \hat{\beta}^{\dagger}_{0}\ket{\Phi}\notag\\
&=\sum_{q_{4}q_{5}q_{6}\neq0}\Big(u_{q_{1}q_{4}}u_{q_{2}q_{5}}v^{*}_{q_{3}q_{6}}w_{q_{4}q_{5}q_{6}}
+v_{q_{1}q_{4}}v_{q_{2}q_{5}}u^{*}_{q_{3}q_{6}}w^{*}_{q_{4}q_{5}q_{6}}\Big),
\end{align}
\end{subequations}
where 
\begin{equation}
\rho^{\frac{3}{2}}_{q_{1}q_{2}}\equiv \bra{\Phi}\hat{\gamma}^{\dagger}_{q_{2}}\hat{\gamma}_{q_{1}} \ket{\Phi}
\simeq \frac{1}{2}\sum_{q_{3}q_{4}}w_{q_{1}q_{3}q_{4}}w^{*}_{q_{2}q_{3}q_{4}}
\end{equation}
 and we approximate
\begin{equation}
(\hat{c}^{\dagger}_{0})^{m}(\hat{c}_{0})^{n}\simeq (N_{0})^{\frac{m+n}{2}} (\hat{\beta}^{\dagger}_{0})^{m}(\hat{\beta}_{0})^{n},
\end{equation}
where $N_{0}$ denotes the number of condensed particles.
Therefore, we obtain an expression for the ground-state energy $\mathcal{E}\equiv \bra{\Phi}\hat{H} \ket{\Phi}$ as
\begin{align}
\mathcal{E}&=\mathcal{E}[\phi_{q_{a}q_{b}},\phi^{*}_{q_{a}q_{b}},w_{q_{a}q_{b}q_{c}},w^{*}_{q_{a}q_{b}q_{c}}]\notag\\
&=\mathcal{E}_{0}+\mathcal{E}_{1}+\mathcal{E}_{\frac{3}{2}}+\mathcal{E}_{2},
\end{align}
where
\begin{subequations}
\begin{align}
\mathcal{E}_{0}=&\bra{\Phi}\hat{H}_{0}\ket{\Phi}=K_{00}N_{0}+\frac{1}{2}U_{00;00}N_{0}^{2},\\
\mathcal{E}_{1}=&\bra{\Phi}\hat{H}_{1}\ket{\Phi}=\sum_{q_{1},q_{2}\neq0}K_{q_{2}q_{1}}\rho_{q_{1}q_{2}}\notag\\
&+N_{0}\sum_{q_{1}q_{2}\neq0}\Big(U_{q_{2}0;q_{1}0}+U_{q_{2}0;0q_{1}}\Big)\rho_{q_{1}q_{2}}\notag\\
&+\frac{N_{0}}{2}\sum_{q_{1}q_{2}\neq0}\Big(U_{00;q_{2}q_{1}}F_{q_{1}q_{2}}+{\rm{C.C.}}\Big),\\
\mathcal{E}_{\frac{3}{2}}=&\bra{\Phi}\hat{H}_{\frac{3}{2}}\ket{\Phi}\notag\\
=&\sqrt{N_{0}}\sum_{q_{1}q_{2}q_{3}\neq0}\Big(U_{0q_{3};q_{2}q_{1}}W_{q_{1}q_{2}q_{3}}+{\rm{C.C.}}\Big),\\
\mathcal{E}_{2}=&\bra{\Phi}\hat{H}_{2}\ket{\Phi}\simeq \frac{1}{2}\sum_{q_{1}q_{2}q_{3}q_{4}\neq0}U_{q_{4}q_{3};q_{2}q_{1}}\notag\\
&\times\Big(F_{q_{1}q_{2}} F^{*}_{q_{3}q_{4}}+\rho_{q_{2}q_{4}}\rho_{q_{1}q_{3}}+\rho_{q_{1}q_{4}}\rho_{q_{2}q_{3}}\Big),
\end{align}
\end{subequations}
where C.C. denotes complex conjugate and we use the decomposition as
\begin{align}
&\bra{\Phi}\hat{c}^{\dagger}_{q_{4}}\hat{c}^{\dagger}_{q_{3}}\hat{c}_{q_{2}}\hat{c}_{q_{1}}\ket{\Phi}\notag\\
&= F^{*}_{q_{4}q_{3}}F_{q_{2}q_{1}}+\rho_{q_{2}q_{4}}\rho_{q_{1}q_{3}}+\rho_{q_{1}q_{4}}\rho_{q_{2}q_{3}}.
\end{align}

In principle, the stationary condition $\delta \mathcal{E}=0$ gives self-consistent equations for $\phi_{q_{1}q_{2}}$ and $w_{q_{1}q_{2}q_{3}}$. 
However, the explicit expression for $\delta\mathcal{E}/\delta\phi^{*}_{q_{a}q_{b}}$ is difficult to obtain unlike the case of homogeneous systems.
This is because $u_{q_{1} q_{2}}$ and $v_{q_{1} q_{2}}$ are expressed by $\phi_{q_{1} q_{2}}$ as given in Eq.\ (\ref{uvphi}), which includes the square root of inverse matrices and is too complicated to perform variational calculations.
For this reason, we introduce a potential $\Omega$ and consider the conditions equivalent to $\delta\mathcal{E}/\delta\phi^{*}_{q_{a}q_{b}}=\delta\mathcal{E}/\delta w^{*}_{q_{a}q_{b}q_{c}}=0$ on the basis of Lagrange multipliers.
Here, we introduce $\Omega$ as
\begin{align}
&\Omega=\mathcal{E}+\mu\left[N-\left(N_{0}+\sum_{q\neq0}\rho_{qq}\right)\right]\notag\\
&+\frac{1}{2}\sum_{q_{1}q_{2}\neq0}\left[\delta_{q_{1}q_{2}}-\sum_{q_{3}\neq0}\left(u^{*}_{q_{3}q_{1}}u_{q_{3}q_{2}}-v_{q_{3}q_{1}}v^{*}_{q_{3}q_{2}}\right) \right]\lambda_{q_{2}q_{1}},
\end{align}
where $\mu$ and $\lambda_{q_{1}q_{2}}$are Lagrange multipliers whose variational conditions give the following constraint conditions:
\begin{subequations}
\begin{align}
&N_{0}+\sum_{q\neq0}\rho_{qq}=N,\label{cons1}\\
&\sum_{q_{3}\neq 0}\Big( u_{q_{1}q_{3}}u^{*}_{q_{2}q_{3}}-v_{q_{1}q_{3}}v^{*}_{q_{2}q_{3}}\Big)=\delta_{q_{1}q_{2}}.\label{cons2}
  \end{align}
 \end{subequations}
Minimizing $\Omega$ instead of $\mathcal{E}$ corresponds to changing the independent variational parameters from $(\phi_{q_{1}q_{2}},w_{q_{1}q_{2}q_{3}},{\rm{C.C.}})$ to $(N_{0},\mu,u_{q_{1}q_{2}},v_{q_{1}q_{2}},\lambda_{q_{1}q_{2}}, w_{q_{1}q_{2}q_{3}},{\rm{C.C.}})$.

Now, we carry out the following variational calculations:
\begin{subequations}
\begin{align}
&\underline{\frac{\delta\Omega}{\delta N_{0}}=0}\notag\\
\to&\mu=\Sigma_{00}-\frac{1}{2}(\Delta_{00}+{\rm{C.C.}}) +\sum_{q_{1}q_{2}\neq0}\Big(U_{00;q_{1}q_{2}}F_{q_{1}q_{2}}+{\rm{C.C.}}\Big)\notag\\
&+\frac{1}{2\sqrt{N_{0}}}\sum_{q_{1}q_{2}q_{3}\neq0}\Big(U_{0q_{3};q_{2}q_{1}}W_{q_{1}q_{2}q_{3}}+{\rm{C.C.}}\Big), {\label{GP0}}\\
&\underline{\frac{\delta\Omega}{\delta u^{*}_{q_{a}q_{b}}}=0 \ \ {\rm{and}} \ \ \frac{\delta\Omega}{\delta v_{q_{a}q_{b}}}=0 } \notag\\
\to&\sum_{q_{1},q_{2}\neq 0}
\begin{bmatrix}
\Sigma_{q_{a}q_{1}}&\Delta_{q_{a}q_{1}} \\
-\Delta^{*}_{q_{a}q_{1}}&-\Sigma^{*}_{q_{a}q_{1}}
\end{bmatrix}
\begin{bmatrix}
u_{q_{1}q_{2}}\\
v^{*}_{q_{1}q_{2}}
\end{bmatrix}
(\delta_{q_{2}q_{b}}+2\rho^{\frac{3}{2}}_{q_{2}q_{b}})\notag\\
&+
\begin{bmatrix}
\chi^{(1)}_{q_{a}q_{b}}\\
\chi^{(2)}_{q_{a}q_{b}}
\end{bmatrix}
=
\sum_{q_{1}\neq 0}\begin{bmatrix}
u_{q_{a}q_{1}}\\
v^{*}_{q_{a}q_{1}}
\end{bmatrix}\lambda_{q_{1}q_{b}},\\
&\underline{\frac{\delta\Omega}{\delta w^{*}_{q_{a}q_{b}q_{c}}}=0}\notag\\
\to& w_{q_{a}q_{b}q_{c}}=-\frac{b_{q_{a}q_{b}q_{c}}}{a_{q_{a}q_{a}}+a_{q_{b}q_{b}}+a_{q_{c}q_{c}}},
 \end{align} \label{GP}
 \end{subequations}
 where we define the following quantities:
 \begin{subequations}
\begin{align}
& \Sigma_{q_{a}q_{b}}\equiv K_{q_{a}q_{b}}+N_{0}(U_{q_{a}0;q_{b}0}+U_{q_{a}0;0q_{b}})\notag\\
 &+\sum_{q_{1}q_{2}\neq0}\Big(U_{q_{a}q_{2};q_{b}q_{1}}+U_{q_{a}q_{2};q_{1}q_{b}}\Big)\rho_{q_{1}q_{2}},\\
 & \Delta_{q_{a}q_{b}}\equiv N_{0}U_{q_{b}q_{a};00}+\sum_{q_{1}q_{2}\neq0}U_{q_{a}q_{b};q_{2}q_{1}}F_{q_{1}q_{2}},\\
&\chi^{(1)}\equiv 2\sqrt{N_{0}}\sum_{q_{1} q_{2} q_{3} q_{4}\neq 0}\Big(U_{0q_{a}; q_{2} q_{1}}v_{q_{1}q_{3}}v_{q_{2} q_{4}}\notag\\
&+ 2U_{q_{a} q_{1}; q_{2}0}u^{*}_{q_{1}q_{3}}v_{q_{2} q_{4}}\Big)w^{*}_{q_{3}  q_{4} q_{b}},\\
& \chi^{(2)}\equiv-2\sqrt{N_{0}}\sum_{q_{1} q_{2} q_{3} q_{4}\neq 0}\Big(U_{q_{1} q_{2}; q_{a}0}u^{*}_{q_{1}q_{3}}u ^{*}_{q_{2} q_{4}}\notag\\
&+ 2U_{0 q_{2}; q_{1}  q_{a}}v_{q_{1}q_{3}} u^{*}_{q_{2} q_{4}}\Big)w^{*}_{q_{3}  q_{4} q_{b}},\\
&a_{q_{a}q_{b}}\equiv \sum_{q_{1}q_{2}\neq 0}\Sigma_{q_{1}q_{2}}(u_{q_{2}q_{b}}u^{*}_{q_{1}q_{a}}+v_{q_{2}q_{b}}v^{*}_{q_{1}q_{a}})\notag\\
&+\sum_{q_{1}q_{2}\neq 0}\Big(\Delta^{*}_{q_{1}q_{2}}u_{q_{2}q_{b}}v_{q_{1}q_{a}}+\Delta_{q_{1}q_{2}}u^{*}_{q_{2}q_{a}}v^{*}_{q_{1}q_{b}}\Big),\\
&b_{q_{a}q_{b}q_{c}}\equiv \sqrt{N_{0}}\sum_{\hat{P}} \hat{P}_{q_{a}q_{b}q_{c}}\sum_{q_{1}q_{2}q_{3}\neq 0}\Bigg[U_{q_{a}q_{b};q_{c}0}u^{*}_{q_{a}q_{1}}u^{*}_{q_{b}q_{2}}v_{q_{c}q_{3}}\notag\\
&+U_{0q_{c};q_{b}q_{a}}v_{q_{a}q_{1}}v_{q_{b}q_{2}}u^{*}_{q_{c}q_{3}}\Bigg]
+\frac{1}{2} \sum_{\hat{P}} \hat{P}_{q_{a}q_{b}q_{c}}\sum_{q_{1}\neq 0,q_{a}}a_{q_{a}q_{1}}w_{q_{1}q_{b}q_{c}}.
 \end{align}
  \end{subequations}

{{

Our theory is based on the premise that we can obtain a set of appropriate one-particle states, that satisfy orthonormality and completeness.
However, we need to consider the real-space deformation of $\varphi_{q}(\bm{r})$ due to the interaction between particles when we carry out numerical calculations.
Although it is difficult to consider it completely, we can discuss the deformation effect focusing only on a condensate wave function.
The deformation can be considered on the basis of the following Gross--Pitaevskii equation including $3/2$-body correlations, which is obtained by $\delta \Omega/\delta \varphi^{*}_{0}(\bm{r})=0$:
\begin{align}
&\int d\bm{r}_{1}\Big[\hat{\mathcal{K}} (\bm{r},\bm{r}_{1})\varphi_{0}(\bm{r}_{1})-\Delta(\bm{r},\bm{r}_{1})\varphi^{*}_{0}(\bm{r}_{1})\Big]\notag\\
= &-\int d\bm{r}_{1}U(|\bm{r}-\bm{r}_{1}|)
\Bigg[ 2F(\bm{r}, \bm{r}_{1})\varphi^{*}_{0}(\bm{r}_{1})+\frac{W( \bm{r},\bm{r}_{1}, \bm{r}_{1})}{\sqrt{N}_{0}}\Bigg], \label{GP1}
\end{align}
where we define the following self-consistent conditions:
\begin{subequations}
\begin{align}
\hat{\mathcal{K}}(\bm{r}_{1},\bm{r}_{2})\equiv &\delta(\bm{r}_{1}-\bm{r}_{2})(\hat{K}_{2}-\mu)+\int d\bm{r}_{3} \Big \{U(\bm{r}_{2}-\bar{\bm{r}}_{3})\notag\\
&\times \big[\rho(\bar{\bm{r}}_{3},\bar{\bm{r}}_{3})+N_{0}|\varphi_{0}(\bar{\bm{r}}_{3})|^{2} \big]\Big \}
+U(\bm{r}_{1}-\bm{r}_{2})
\notag\\
&\times \big[\rho(\bm{r}_{1},\bm{r}_{2})
+N_{0}\varphi_{0}(\bm{r}_{1})\varphi^{*}_{0}(\bm{r}_{2}) \big]\notag\\
&\equiv\delta(\bm{r}_{1}-\bm{r}_{2})(\hat{K}_{2}-\mu)+\Sigma(\bm{r}_{1},\bm{r}_{2}) \label{HF}\\
\Delta(\bm{r}_{1},\bm{r}_{2})\equiv& U(\bm{r}_{1}-\bm{r}_{2})\big[F(\bm{r}_{1},\bm{r}_{2})+N_{0}\varphi_{0}(\bm{r}_{1})\varphi_{0}(\bm{r}_{2})\big],\label{Pair}\\
\rho (\bm{r}_{1},\bm{r}_{2})\equiv & \sum_{q_{1}q_{2}\neq 0} \rho_{q_{1}q_{2}}\varphi_{q_{1}}(\bm{r}_{1})\varphi^{*}_{q_{2}}(\bm{r}_{2}), \\
F (\bm{r}_{1},\bm{r}_{2})\equiv & \sum_{q_{1}q_{2}\neq 0} F_{q_{1}q_{2}}\varphi_{q_{1}}(\bm{r}_{1})\varphi_{q_{2}}(\bm{r}_{2}), \\
W (\bm{r}_{1},\bm{r}_{2},\bm{r}_{3})\equiv &\sum_{q_{1}q_{2}q_{3}\neq 0} W_{q_{1}q_{2}q_{3}}\varphi_{q_{1}}(\bm{r}_{1})\varphi_{q_{2}}(\bm{r}_{2})\varphi^{*}_{q_{3}}(\bm{r}_{3}).
\end{align}
\end{subequations}
By solving Eq.\ (\ref{GP1}) with the self-consistent conditions, we obtain $\varphi_{0}(\bm{r})$ deformed by interactions between particles.
However, the Gross--Pitaevskii equation does not guarantee the orthogonality of one-particle states, i.e., it is necessary to check the condition $\braket{0|q}=\delta_{0q}$ for $q>0$.
To evaluate all the contributions from $\ket{q}$ strictly and appropriately, we need to obtain $\varphi_{0}(\bm{r})$ and $\varphi_{q}(\bm{r})$ self-consistently with a constraint condition $\braket{q|q'}=\delta_{qq'}$.
Although it remains as a future technical task, $\varphi_{q}(\bm{r})$ may be obtained from the stationary condition $\delta\Omega/\delta \varphi^{*}_{q}(\bm{r})=0$ in the same manner as $\varphi_{0}(\bm{r})$.
}}

{{This formulation can be applied to the homogeneous BEC\cite{kita3/2} by changing the subscript as $q\to\bm{k}$ and the basis function as $\varphi_{q}(\bm{r})\to e^{i\bm{k}\cdot \bm{r}}/\sqrt{\mathcal{V}}$, where $\bm{k}$ denotes the wave number of a particle.}}
In this case, the basic matrix elements are given by
\begin{align} 
&K_{\bm{k}\bm{k}'}=\delta_{\bm{k},\bm{k}'}\varepsilon_{\bm{k}}, \ U_{\bm{k}_{1}\bm{k}_{2};\bm{k}_{3}\bm{k}_{4}}=\frac{\delta_{\bm{k}_{1}+\bm{k}_{2},\bm{k}_{3}+\bm{k}_{4}}}{\mathcal{V}}U_{|\bm{k}_{1}-\bm{k}_{3}|},\notag\\
&\phi_{\bm{k}\bm{k}'}=\delta_{\bm{k},-\bm{k}'}\phi_{\bm{k}}, 
\end{align} 
with $\phi_{\bm{k}}=\phi_{-\bm{k}}$ and 
\begin{align} 
\varepsilon_{\bm{k}}=\frac{\hbar^{2}|\bm{k}|^{2}}{2m} , \ U_{|\bm{k}|}=\int d\bm{r} U(|\bm{r}|)e^{-i\bm{k}\cdot \bm{r}}.
\end{align} 

\section{Application to a System Trapped by a One-Dimensional Harmonic Oscillator}
As one of the simplest numerical examples, we consider a 1D system ($\bm{r}\to z$) trapped by a harmonic oscillator $V(z)=m\omega^{2}z^{2}/2$ with a short-range contact potential written as $U(|z_{a}-z_{b}|)=g\delta(z_{a}-z_{b})$.{{
Using the potential, we transform Eq.\ (\ref{GP1}) into
\begin{align}
&\Bigg\{-\frac{\hbar^{2}}{2m}\frac{\partial^{2}}{\partial z ^{2}} +\frac{m\omega^{2}z^{2}}{2}-\mu+\Sigma(z) \Bigg\} \varphi_{0}(z)-\Delta(z) \varphi^{*}_{0}(z)\notag\\
= &-g\Bigg[ 2F(z)\varphi^{*}_{0}(z)+\frac{W(z)}{\sqrt{N_{0}}}\Bigg], \label{GP2}
\end{align}
where $X(\bm{r})= X(\bm{r}, \bm{r})$ ($X=\Sigma$, $\Delta$, and $F$) and $W(\bm{r})= W(\bm{r}, \bm{r}, \bm{r})$. 
In the following calculation, we set the units of energy $\varepsilon_{\omega}=\hbar\omega/2$ and length $l_{\omega}=(\hbar/m\omega)^{\frac{1}{2}}$.}}

{{We point out that it is crucial to choose an appropriate $\ket{q}$ that corresponds to the external potential considered in the present formulation.
In the limit $g\to0$, the condensate wave function $\varphi_{0}(\bm{r})$ becomes a Gaussian.
On the other hand, {{$\varphi_{0}(\bm{r})$}} deforms due to the nonlinear term of the Gross--Pitaevskii equation when we set a larger $g$, such as in the case of a Thomas--Fermi BEC\cite{Pethicsmith}.
With these considerations, we propose two approximations in the weak-coupling region as follows: 
\begin{itemize}
\item[(i)]: $\ket{q}\simeq \ket{n}$ for all $n\geq 0$,
\item[(ii)]: \ $\braket{\bm{r}|0} =\varphi^{\rm{GP}}_{0}(\bm{r})$ and $\ket{q}\simeq \ket{n}$ for $n>0$,
\end{itemize}
where $\varphi^{\rm{GP}}_{0}(\bm{r})$ represents the solution of Eq.\ (\ref{GP2}); integer $n\leq n_{\rm{cut}}$ is a quantum number that characterizes the energy levels of a harmonic oscillator and $n_{\rm{cut}}$ is the cutoff energy level.}}
On the basis of the approximation, $\varphi_{n}(z)=\braket{z|n}$ is given as
\begin{equation}
\varphi_{n}(z)\equiv \left(\frac{1}{2^{n}n! \sqrt{\pi}l_{\omega}}\right)^{\frac{1}{2}}H_{n}\left(\frac{z}{l_{\omega}} \right){\rm{exp}}\left(-\frac{z^{2}}{2l^{2}_{\omega}} \right),
\end{equation}
where $H_{n}$ denotes the $n$th Hermite polynomial.
Hence, we obtain $K_{n_{1}n_{2}}$ and $U_{n_{1}n_{2};n_{3}n_{4}}$ as
\begin{subequations}
\begin{align}
&K_{n_{1}n_{2}}=\delta_{n_{1}n_{2}}(2n_{1}+1)\varepsilon_{\omega},\\
&U_{n_{1}n_{2};n_{3}n_{4}}= g\int_{-l_{\rm{cut}}}^{l_{\rm{cut}}} dz \varphi_{n_{1}}(z)\varphi_{n_{2}}(z)\varphi_{n_{3}}(z)\varphi_{n_{4}}(z),
\end{align}
\end{subequations}
where $l_{\rm{cut}}$ is the cutoff length for numerical calculations.

To carry out the numerical calculations, we introduce the external parameter (coupling constant) as
\begin{equation}
\alpha\equiv \frac{mg l_{\omega}}{\hbar^{2}}=\frac{1}{2}\frac{g}{\varepsilon_{\omega} l_{\omega}} \ll 1.\label{charact2}
\end{equation}
This parameter denotes the ratio of the scales for the correlation of particles and the harmonic oscillator potential. 
In this work, we carry out the numerical calculations for $N=1000$ and $\alpha\sim 1.0\times 10^{-3}$, where the BEC has an approximately Gaussian profile\cite{Petrov}.
{{In addition, we neglect the $O(|W|^{2})$ terms in the self-consistent conditions because they give only a small correction to the ground-state wave function in the weak-coupling regime.}}
We choose $n_{\rm{cut}}=40$ ($\varepsilon_{n_{\rm{cut}}}=81\varepsilon_{\omega}$) and $l_{\rm{cut}}=10l_{\omega}$ ($m\omega l_{\omega}^{2}/2=100\varepsilon_{\omega} \sim \varepsilon_{n_{\rm{cut}}}$) for the numerical calculations.
We start the initial self-consistent calculation by substituting the trivial solutions for $g=0$ and renew the solutions one after another while mixing the old and new solutions with the weight ratio of $80$ : $20$.

Now, we discuss the numerical results.
First, we show the ground-state energy $\mathcal{E}_{\sigma}$ ($\sigma=0,1,3/2,2$) to explain the respective energy scales {{in Tables \ref{GE1}} and \ref{GE2}}.
From these tables, we see that $|\mathcal{E}_{0}|\gg|\mathcal{E}_{1}|\gg |\mathcal{E}_{2}|>|\mathcal{E}_{\frac{3}{2}}|$.
However, $\mathcal{E}_{\frac{3}{2}}$ is comparable to $\mathcal{E}_{2}$ around $\alpha\sim 1.0\times 10^{-3}$.
In addition, $|\mathcal{E}_{\frac{3}{2}}|/|\mathcal{E}_{2}|$ seems to increase monotonically as a function of $\alpha$ so that the $3/2$-body correlation may be dominant in the relatively strong  coupling system, such as the Thomas--Fermi BEC regime.
{{Comparing Tables \ref{GE1} and \ref{GE2}, we find that approximation (ii) yields lower total ground-state energies than approximation (i) because the deformation of the condensate wave function also lowers the ground-state energy.
Hence, one might conclude that (ii) is better than (i).
However, (ii) appears to break the orthogonality relation, i.e., $\braket{0|n}\neq \delta_{0n}$ for $n>0$.
To evaluate the ground-state energies more quantitatively, setting appropriate one-particle states with orthogonality relations remains a future task.
}}

Incorporating more variational parameters in the theory is expected to yield a better estimate for the ground-state energy.
To see this explicitly, we perform our variational calculations on the basis of the following ground states:
\begin{itemize}
\item[(1)]$\ket{\Phi_{\rm{GP}}}$: We set $\phi_{n_{1}n_{2}}=w_{n_{1}n_{2}n_{3}}=0$.
\item[(2)]$\ket{\Phi_{\rm{Bog}}}$: We obtain $\phi_{n_{1}n_{2}}$ while fixing $\rho_{n_{1}n_{2}}=F_{n_{1}n_{2}}=w_{n_{1}n_{2}n_{3}}=0$.
\item[(3)] $\ket{\Phi_{\rm{HFB}}}$: We set $\lambda_{n_{1}n_{2}}=E_{n_{1}}\delta_{n_{1}n_{2}}$ and $w_{n_{1}n_{2}n_{3}}=0$. This is equivalent to the problem of diagonalizing the effective Hamiltonian called HFB theory\cite{HFB}.
\item[(4)]$\ket{\Phi_{\rm{GA}}}$: All the variational parameters except $w_{n_{1}n_{2}n_{3}}$ are calculated self-consistently. The ground state is equivalent to the GA wave function.
\item[(5)]$\ket{\Phi}$: All the variational parameters are calculated self-consistently. 
\end{itemize}
Using the ground states, we evaluate the energy differences defined by
 \begin{subequations}
 \begin{align}
& \Delta \mathcal{E}_{I}\equiv  \bra{\Phi} \hat{H}\ket{\Phi}-\bra{\Phi_{\rm{GP}}} \hat{H}\ket{\Phi_{\rm{GP}}},\\
& \Delta \mathcal{E}_{I\hspace{-.1em}I}\equiv  \bra{\Phi_{\rm{GA}}} \hat{H}\ket{\Phi_{\rm{GA}}}-\bra{\Phi_{\rm{Bog}}} \hat{H}\ket{\Phi_{\rm{Bog}}} ,\\
& \Delta \mathcal{E}_{I\hspace{-.1em}I\hspace{-.1em}I}\equiv  \bra{\Phi} \hat{H}\ket{\Phi}-\bra{\Phi_{\rm{HFB}}} \hat{H}\ket{\Phi_{\rm{HFB}}} ,\\
&\Delta \mathcal{E}_{I\hspace{-.1em}V}\equiv  \bra{\Phi} \hat{H}\ket{\Phi}-\bra{\Phi_{\rm{GA}}} \hat{H}\ket{\Phi_{\rm{GA}}}.
 \end{align}
 \end{subequations}
{{From Tables \ref{dGE1} and \ref{dGE2}}}, we subsequently see the relation $\bra{\Phi} \hat{H}\ket{\Phi}<\bra{\Phi_{\rm{GA}}} \hat{H}\ket{\Phi_{\rm{GA}}}<\bra{\Phi_{\rm{HFB}}} \hat{H}\ket{\Phi_{\rm{HFB}}}<\bra{\Phi_{\rm{Bog}}} \hat{H}\ket{\Phi_{\rm{Bog}}}<\bra{\Phi_{\rm{GP}}} \hat{H}\ket{\Phi_{\rm{GP}}}$; thus, $\ket{\Phi}$ seems to be the best solution in terms of constructing the variational wave function.
The reason why $\ket{\Phi_{\rm{GA}}}$ gives lower energy than $\ket{\Phi_{\rm{HFB}}}$ is traced back to the difference in the manner of setting $\underline{\lambda}$, i.e., the difference between $\underline{u}\ \underline{u}^{\dagger}-\underline{v}\ \underline{v}^{\dagger}=\underline{1}$ for $\ket{\Phi_{\rm{GA}}}$ and ${\rm{Tr}}[\underline{u}\ \underline{u}^{\dagger}-\underline{v}\ \underline{v}^{\dagger}]=1$ for $\ket{\Phi_{\rm{HFB}}}$. In the latter case, the quasiparticles do not satisfy the Bose commutator relations because the condition for the off-diagonal parts of $(\underline{u}\ \underline{u}^{\dagger}-\underline{v}\ \underline{v}^{\dagger})$ is not considered to be appropriate. In contrast, $\ket{\Phi_{\rm{GA}}}$ with $\underline{u}$ and $\underline{v}$ satisfying all the conditions of Eq.\ (\ref{uv}) gives a lower ground-state energy than $\ket{\Phi_{\rm{HFB}}}$.
Thus, the appropriate consideration for commutator relations of quasiparticles is indispensable for obtaining the lower ground-state energy.

\begin{table}[t]
{{
\caption{Ground-state energies based on approximation (i) with $\tilde{\mathcal{E}}_{\sigma}=\mathcal{E}_{\sigma}/(N\varepsilon_{\omega})$ and $\tilde{\alpha}=\alpha\times 10^{3}$ .}
\centering
\begin{tabular}{cccccc}
\hline\\[-2.0ex]
$\tilde{\alpha}$ &$\tilde{\mathcal{E}}_{0}$ &$\tilde{\mathcal{E}}_{1}$&$\tilde{\mathcal{E}}_{\frac{3}{2}}$& $\tilde{\mathcal{E}}_{2}$ &$\tilde{\mathcal{E}}_{\frac{3}{2}}/\tilde{\mathcal{E}}_{2}$ \\ [0.5ex] 
\hline\hline
$0.1$&$ 1.0399 $&$ -5.49 \times 10^{-7} $&$ -6.51 \times 10^{-13}  $&$ 1.64 \times 10^{-11} $ & -0.040\\[0.2ex] 
$0.5$&$ 1.20 $&$ -1.22\times 10^{-5}  $&$ -3.60\times 10^{-10} $&$ 1.93\times 10^{-9}$ &-0.19 \\[0.2ex]
$1.5$&$1.60 $&$ -7.85\times 10^{-5} $&$ -2.36\times 10^{-8} $&$4.85\times 10^{-8} $&-0.49\\[0.2ex]
$2.5$&$ 2.00 $&$ -1.22\times 10^{-4} $&$ -1.69\times 10^{-7} $&$ 2.34\times 10^{-7} $ & -0.72\\[0.2ex]
\hline
\end{tabular}
\label{GE1}

\caption{Ground-state energies based on approximation (ii) with $\tilde{\mathcal{E}}_{\sigma}=\mathcal{E}_{\sigma}/(N\varepsilon_{\omega})$ and $\tilde{\alpha}=\alpha\times 10^{3}$.}
\centering
\begin{tabular}{cccccc}
\hline\\[-2.0ex]
$\tilde{\alpha}$ &$\tilde{\mathcal{E}}_{0}$ &$\tilde{\mathcal{E}}_{1}$&$\tilde{\mathcal{E}}_{\frac{3}{2}}$& $\tilde{\mathcal{E}}_{2}$ &$\tilde{\mathcal{E}}_{\frac{3}{2}}/\tilde{\mathcal{E}}_{2}$ \\ [0.5ex] 
\hline\hline
$0.1$&$ 1.0397 $&$ -5.47 \times 10^{-7}$&$ -6.57\times 10^{-13} $&$ 1.64\times 10^{-11}$&-0.040 \\[0.2ex] 
$0.5$&$ 1.19 $&$ -1.19 \times 10^{-5}$&$ -3.72\times 10^{-10} $&$ 1.94\times 10^{-9} $&-0.19 \\[0.2ex]
$1.5$&$ 1.56 $&$ -7.71 \times 10^{-5}$&$ -2.43\times 10^{-8} $&$ 4.73\times 10^{-8}$&-0.51 \\[0.2ex]
$2.5$&$ 1.89 $&$ -1.55 \times 10^{-4}$&$ -1.57\times 10^{-7} $&$ 2.04\times 10^{-7} $&-0.77 \\[0.2ex]
\hline
\end{tabular}
\label{GE2}

\caption{Energy differences based on approximation (i) with  $\Delta\tilde{\mathcal{E}}_{\sigma}=\Delta\mathcal{E}_{\sigma}/(N\varepsilon_{\omega})$ and $\tilde{\alpha}=\alpha\times 10^{3}$.}
\centering
\begin{tabular}{ccccc}
\hline\\[-2.0ex]
$\tilde{\alpha}$&  $\Delta\tilde{\mathcal{E}}_{I}$ &$\Delta\tilde{\mathcal{E}}_{I\hspace{-.1em}I}$ &$\Delta\tilde{\mathcal{E}}_{I\hspace{-.1em}I\hspace{-.1em}I}$&$\Delta\tilde{\mathcal{E}}_{I\hspace{-.1em}V}$ \\ [0.5ex]
\hline\hline
$0.1$&$-7.26 \times 10^{-7}$&$-4.44 \times10^{-16}$&$  -6.59 \times10^{-13}$&$  -6.52 \times 10^{-13} $\\[0.2ex] 
$0.5$&$ -1.75\times 10^{-5} $&$-2.81\times 10^{-13}$&$ -4.46\times 10^{-10} $&$-3.60\times 10^{-10} $\\[0.2ex]
$1.5$&$ -1.46\times 10^{-4}$&$-2.83\times 10^{-11}$&$ -6.23\times 10^{-8}$&$-2.36\times 10^{-8} $\\[0.2ex]
$2.5$&$-3.87\times 10^{-4} $&$ -3.45\times 10^{-10}$&$ -7.94\times 10^{-7} $&$ -1.69\times 10^{-7} $\\[0.2ex]
\hline
\end{tabular}
\label{dGE1}

\caption{Energy differences based on approximation (ii) with $\Delta\tilde{\mathcal{E}}_{\sigma}=\Delta\mathcal{E}_{\sigma}/(N\varepsilon_{\omega})$ and $\tilde{\alpha}=\alpha\times 10^{3}$.}
\centering
\begin{tabular}{ccccc}
\hline\\[-2.0ex]
$\tilde{\alpha}$&  $\Delta\tilde{\mathcal{E}}_{I}$ &$\Delta\tilde{\mathcal{E}}_{I\hspace{-.1em}I}$ &$\Delta\tilde{\mathcal{E}}_{I\hspace{-.1em}I\hspace{-.1em}I}$&$\Delta\tilde{\mathcal{E}}_{I\hspace{-.1em}V}$ \\ [0.5ex]
\hline\hline
$0.1$&$ -7.24\times 10^{-7} $&$ -1.22\times 10^{-14} $&$-6.63\times 10^{-13} $&$ -6.56\times 10^{-13} $\\[0.2ex] 
$0.5$&$ -1.72\times 10^{-5} $&$ -5.61\times 10^{-12} $&$-4.61\times 10^{-10} $&$ -3.72\times 10^{-10}  $\\[0.2ex]
$1.5$&$ -1.39\times 10^{-4} $&$ -2.77\times 10^{-10} $&$-6.27\times 10^{-8} $&$ -2.43\times 10^{-8}  $\\[0.2ex]
$2.5$&$ -3.51\times10^{-4} $&$ -1.40\times 10^{-9} $&$-6.58\times 10^{-7} $&$ -1.57\times10^{-7} $\\[0.2ex]
\hline
\end{tabular}

\label{dGE2}
}}
\end{table}

In addition, $|\Delta \mathcal{E}_{I\hspace{-.1em}V}|$ is roughly $100-1000$ times larger than $|\Delta \mathcal{E}_{I\hspace{-.1em}I}|$.
{{This result indicates that the $3/2$-body correlations contribute to the decrease in the ground-state energies more than the $2$-body correlations.}}
These results agree with the results of homogeneous systems\cite{kita3/2,WK}.
Therefore, the mean-field approximation for inhomogeneous BECs characterized by the discretized energy levels may not be effective quantitatively even in the weak-coupling region, similarly to the homogeneous systems.

{{
Finally, we discuss what is qualitatively peculiar to the trapped systems concerning the $3/2$-body processes on the ground state around $\alpha\sim 1.0\times 10^{-3}$.
To do this, we introduce the energy density as
\begin{equation}
\mathcal{E}_{i}=\int_{-l_{\rm{cut}}}^{l_{\rm{cut}}} d\bm{z}\eta _{i}(z), 
\end{equation}
where $i=\frac{3}{2}$ or $2$.
Using this quantity, we can present a local discussion of the correlation energy of inhomogeneous systems.
Figure. \ref{fig1} shows the spatial dependence of energy densities $\eta_{\frac{3}{2}}(z)$, $\eta_{2}(z)$, and $\eta_{\frac{3}{2}}(z)+\eta_{2}(z)$ for $\alpha=2.5\times 10^{-3}$ within approximation (i).
As shown in the figure, both $\eta_{\frac{3}{2}}(z)<0$ and $\eta_{2}(z)>0$ have large off-center peaks around $z\sim 0.9l_{\omega} $. Consequently, $\eta_{\frac{3}{2}}(z)+\eta_{2}(z)<\eta_{2}(z)$ becomes more spatially homogeneous than $\eta_{2} (z) $. 
Assuming $\mathcal{E}_{\frac{3}{2}}+ \mathcal{E}_{{2}}$ is the effective correlation energy, the $3/2$-body correlations play the roles of (1) suppressing the effective correlation energy and (2) shaping the effective correlation-energy density more homogeneously.
Note that approximation (ii) qualitatively yields the same result as (i).

\begin{figure}[t]
   {{     \begin{center}
                \includegraphics[width=0.9\linewidth]{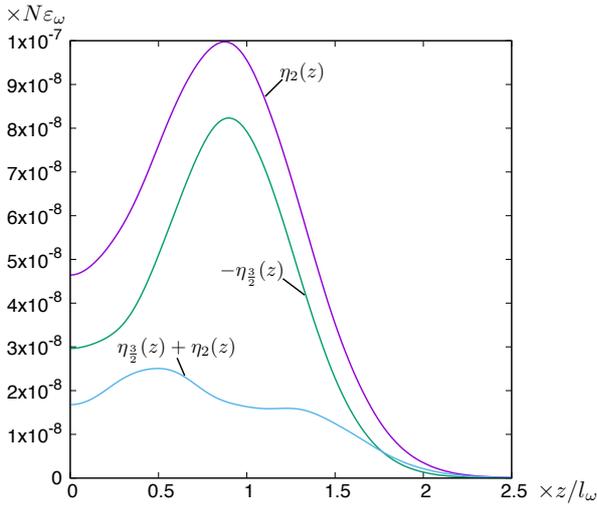}
                \end{center}
\caption{(Color online) Energy densities $\eta_{\frac{3}{2}}(z)$, $\eta_{2}(z)$, and $\eta_{\frac{3}{2}}(z)+\eta_{2}(z)$ for $\alpha=2.5\times 10^{-3}$ using approximation (i).
} 
\label{fig1}}}
\end{figure}

}}

\section{Summary and Conclusion}
We have constructed the variational wave function for an inhomogeneous BEC including not only the mean-field $2$-body correlations but also the $3/2$-body correlations beyond the mean-field approximation.
Using the variational wave function, we have carried out a numerical calculation to evaluate the ground-state energy of a 1D BEC trapped by a harmonic oscillator.
Our numerical result shows that $3/2$-body correlations decrease the ground-state energies even in a trapped system characterized by the discretized energy level, and their contributions are comparable to those of $2$-body correlations, which agree with the results of homogeneous cases\cite{kita3/2,WK}.
Therefore, when we consider the contributions from noncondensates, self-consistent mean-field approximations may not be valid in BEC systems and $3/2$-body correlations should be incorporated.
This wave function is expected to give physical pictures beyond mean-field contributions in inhomogeneous systems more microscopically.

{{
\section*{Acknowledgements}
W. K. is a JSPS Research Fellow, and this work was supported in part by JSPS KAKENHI Grant Number 18J13241.}}

\end{document}